\documentclass[aps,prd,twocolumn,showpacs, superscriptaddress,amsmath,amssymb,nofootinbib]{revtex4-1}

\usepackage[pdftex]{graphicx}
\DeclareGraphicsExtensions{.pdf}
\usepackage{wasysym}
\usepackage{color}
\usepackage{enumerate}

\begin{document}

%\title{Is there a kinematic cutoff in the production of electronic excitation from sub-keV nuclear recoils?}
\title{Atomic limits in the search for galactic dark matter}

\author{Peter~Sorensen}
\email{pfsorensen@lbl.gov}
\affiliation{Lawrence Berkeley National Laboratory, 1 Cyclotron Rd., Berkeley, CA 94720, USA}

\date{\today}
\begin{abstract}
Direct searches for low mass dark matter particles via scattering off target nuclei require detection of recoiling atoms with energies of $\sim1$~keV or less. The amount of electronic excitation produced by such atoms is quenched relative to a recoiling electron of the same energy. The Lindhard model of this quenching, as originally formulated, remains widely used after more than 50 years. The present work shows that for very small energies, a simplifying approximation of that model must be removed. Implications for the sensitivity of direct detection experiments are discussed.
\end{abstract}
%\pacs{95.35.+d, 95.55.Vj, 14.80.Nb, 29.40.-n }% PACS, the Physics and Astronomy

\maketitle

\section{Introduction} \label{sec1}
The possibility of low-mass (usually: $m\lesssim10$~GeV) cold dark matter candidates is theoretically interesting (see {\it e.g.} \cite{Hooper:2012cw,Lin:2011gj}) and experimentally challenging (see {\it e.g.} \cite{Pyle:2012mz,Angle:2011th}). The experimental signature for direct detection generally reduces to detection of recoiling target atoms (``nuclear recoils''), following a scattering event. Recently, interest in dark matter coupling to electrons has also increased \cite{Essig:2011nj,Graham:2012su}. However, the present work focuses on the former situation. For particles bound in a galactic dark matter halo, nuclear recoil energies are typically $\mathcal{O}$(keV) or less, due to the galactic escape velocity $v\simeq 0.002c$. The amount of electronic excitation produced by a recoiling atom is quenched by approximately $\times5$ or more relative to a recoiling electron of the same energy.  Arguably the best theoretical prediction of this quenching is given by Lindhard {\it et al.} \cite{Lindhard:1963a}. Most measured values show good agreement in germanium and silicon, and decent agreement in argon and xenon. 

Experiments such as CoGeNT \cite{Aalseth:2012if} and CDMSlite \cite{Agnese:2013jaa} have performed optimized, dedicated searches for low mass dark matter. The latter uses the Lindhard quenching model parameterization to reconstruct the nuclear recoil energy of events, while the former uses a slightly more optimistic variation of the basic model. In both cases it has been reasonably argued that data support the choice, and in both cases the energy threshold for nuclear recoils is $\sim1$~keV. 

On the other hand, experiments such as XENON10 \cite{Angle:2011th} and DAMIC \cite{Barreto:2011zu} have estimated their sensitivity to elastic dark matter scattering using an extrapolation of the Lindhard quenching prediction to reconstruct the nuclear recoil energy. The necessity of this approach has arisen from to the combination of single (or nearly single) electron detection thresholds along with a complete lack of nuclear recoil quenching data for such small energies. The G2 experiments LZ and SuperCDMS SNOLAB face a similar situation in projecting their sensitivity \cite{Cushman:2013zza}.

At the same time, it is anecdotally known that the validity of the Lindhard model in any material is questionable at very low energies. This can probably be traced to the cautionary statement appearing in \cite{Lindhard:1963b} that ``at extremely low $\varepsilon$-values\footnote{$\varepsilon$ is a reduced energy defined by Eq. \ref{eq:epsilon}} , $\varepsilon\lesssim10^{-2}$, the nuclear scattering and stopping becomes somewhat uncertain, because the Thomas-Fermi treatment is a crude approximation when the ion and the atom do not come close to each other.''

%``...the theory is somewhat uncertain at quite low $\varepsilon$-values, i.e. $\varepsilon\lesssim10^{-2}$.'' 
The basic problem, then, is that the requisite data for low energy nuclear recoils are sparse to nonexistent. With the exception of germanium, reconstructed energies smaller than a few keV must rely on a model. And, the most widely used model is most uncertain in this regime. 

This article will examine the sources of this uncertainty, undo a simple approximation of the original treatment, and obtain a new solution of the original model in the energy range of interest for low-mass dark matter, i.e. $\varepsilon\lesssim10^{-2}$ (generally, $E$ less than a few keV). It is worthwhile to begin with a very brief summary of the ``admittedly elaborate'' original treatment \cite{Lindhard:1963a,Lindhard:1963b,Lindhard:1968}. 
% note also: http://physics.nist.gov/PhysRefData/Star/Text/programs.html

\section{A very brief summary of the Lindhard Model}
The discussion in this section gives a broad-brush picture of the  steps leading to the result known to the dark matter direct detection community as the Lindhard Model. It owes much to \cite{Wilson:1977,Ziegler:1985}, however, the notation follows that of \cite{Lindhard:1963a}. Most formulae and their derivations are intentionally left to the references. 

For a recoiling atom of energy E, what portion $\eta$ of the total energy loss is given to electrons? The remainder of the energy loss $\nu$ is assumed to be given to atomic motion, viz. $\eta + \nu = E$. Unless phonon energy is measured, $\eta$ is an upper limit to the available signal in a particle detector. For simplicity, fluctuations are treated separately and the model is written in terms of average quantities $\bar{\eta}$ and $\bar{\nu}$. An additional simplification is obtained in the case that the projectile and target atoms have the same atomic and mass numbers. The present work therefore focuses on four materials which satisfy this criteria, and are of current interest to the direct detection of dark matter: germanium, silicon, xenon and argon.
 
A recoiling atom of any appreciable energy will undergo a cascade of collisions in its slowing down. Thus, for a given energy $E$, the competition between electronic and nuclear cross sections {\it at all smaller energies} contribute to the partitioning between $\bar{\eta}$ and $\bar{\nu}$ at the energy $E$. The average energy given to atomic motion is obtained by integrating over all possibilities. This physical picture is described by
\begin{equation} \label{eq:idiff}
k \varepsilon^{1/2} \bar{\nu}'(\varepsilon) = \int_{0}^{\varepsilon^2} \frac{dt}{2t^{3/2}} f(t^{1/2}) \left\{ \bar{\nu}(\varepsilon-\frac{t}{\varepsilon}) -\bar{\nu}(\varepsilon) + \bar{\nu}(\frac{t}{\varepsilon}) \right\},
\end{equation} 
in which
\begin{equation} \label{eq:epsilon}
\varepsilon \equiv  E \frac{a}{2Z^2e^2},
\end{equation}
is a dimensionless reduced energy. The details of the other symbols in this equation are explained presently in Sec. \ref{ss:nucl} and Sec. \ref{ss:elec}. The three terms in curly braces refer to the energy of the target atom after a collision, the projectile atom before a collision and the projectile atom after a collision. 

Four key approximations underpin this equation:
\begin{enumerate}[(A)]
\item Ionized electrons do not produce recoil atoms of appreciable energy.
\item The atomic binding energy $u$ of electrons is negligible.
\item Energy transfers to electrons are small relative to energy transfers to atoms.
%\item Energy transfers to electrons $T_{e}$ are small relative to energy transfers to atoms $T_{n}$. %This is reasonable because $T_{e} \sim m_e/m_t$. 
\item The treatment of atomic and electronic collisions are separable.
\end{enumerate}
Of the four, approximation (B) is the most obviously troubling in the limit of low energy recoils.

\subsection{Nuclear stopping cross section} \label{ss:nucl}
The nuclear scattering is modeled as two-body scattering in a screened Coulomb potential, $V(r) = (e^2Z^2/r) \phi_0(r/a)$. The function $\phi_0(r/a)$ is a single atom Thomas-Fermi screening function with length scale $a=0.8853a_0/Z^{1/3}$. In this equation, $a_0$ is the Bohr radius. The standard technique is to extend this screening function to a pair of atoms via a suitable scaling of $a$. Lindhard used $a=0.8853a_0/(Z^{1/3} \sqrt{2})$, though other slightly different scalings have been argued. 

Classical Mechanics then allows a further simplification into that of a single particle moving under a central potential \footnote{I emphasize this elementary step because of its tacit assumption of spherical symmetry $-$ which is of course broken by e.g. polarization of the medium.}. This can be solved for the orbit equation for two-body central force scattering, which gives the scattering angle $\Theta$ (in center-of-mass coordinates) in terms of the initial particle energy and the impact parameter. The nuclear stopping power $S_n(\varepsilon)$ is just the average energy transfer, integrating over all possible impact parameters.

The trick realized by Lindhard et al. was a change of variables, defining $t=\varepsilon^2~\mbox{sin}^2(\Theta/2)$. The nuclear stopping is then defined by an integral over a function $f(t^{1/2})$ ({\it cf.} Fig. 2 of \cite{Lindhard:1963a}). To emphasize, what was previously a function of three variables ($\Theta$, impact parameter and initial particle energy) is now a function of a single variable, $t$. Considering that these simplifications lead to the first solutions of projectile range and energy loss within a single model, we can perhaps forgive the authors for referring to their results as the ``magic formula'' \cite{Lindhard:1968}.

%\begin{equation}
%f(t^{1/2}) = \frac{2t^{3/2}}{\pi a^2} \frac{d\sigma}{dt},
%\end{equation}
% is proportional to the both the energy $\varepsilon$ and the energy transfer $T=\mbox{E}~\mbox{sin}^2(\Theta/2)$. 

\subsection{Electronic stopping cross section} \label{ss:elec}
The electronic stopping power $S_e(\varepsilon)$ can be written as 
\begin{equation} \label{eq:dedr}
d\varepsilon/d\rho=k\varepsilon^{1/2},
\end{equation}
 where $\rho$ is a reduced range. Velocity proportional stopping is a very generic prediction in most models of electronic stopping power. However, calculations of the slope $k$ vary by up to a factor $\times2$ or more \cite{Land:1977}. In this article, I follow the calculation of \cite{Lindhard:1963a} (unless noted otherwise), using  
\begin{equation} \label{eq:k}
 k=0.133Z^{2/3}A^{-1/2},
\end{equation} 
where $A$ is the mass number of the material. 

Eq.~\ref{eq:dedr} has been clearly verified for the simplest case of antiproton stopping \cite{Moller:2002}. When one looks at a wider array of electronic stopping power data for very slow heavy ions \cite{Sigmund:2015}, it is clear that velocity proportionality is generally observed. 

A sticking point for the direct detection community is that most of these models treat atomic electrons as an electron gas. Since our detector targets tend to be semiconductors or large band-gap insulators rather than metals, one might expect a deviation from velocity proportionality. Interestingly, velocity proportionality is still observed \cite{Markin:2009} in materials with large band gaps. However, a non-zero $\varepsilon$-intercept (in the sense of Eq. \ref{eq:dedr}) seen in some data indicates the presence of a threshold velocity, below which the projectile suffers no electronic energy loss.

For point-like projectiles, this threshold velocity is calculable from simple kinematic constraints \cite{Ahlen:1983}. For atomic projectiles, these arguments $-$ which were applied in \cite{Collar:2010gg} $-$ are not applicable. Yet, intuition suggests that the size of a material's band gap should directly affect the low-energy electronic response to nuclear recoils. Within the context of the Lindhard model, the connection lies in approximation (B).

To emphasize, velocity proportional stopping is well established, but strict velocity proportional stopping in the sense of Eq. \ref{eq:dedr}, with an $\varepsilon$-intercept of zero, is not well established. Nevertheless, I follow the assumption of the original work \cite{Lindhard:1963a} and assume Eq. \ref{eq:dedr} holds in the limit $\varepsilon \rightarrow 0$.
  
\subsection{The standard solution}
Analytical solution of Eq. \ref{eq:idiff} is possible only for unrealistic (unscreened) atomic potentials. Lindhard {\it et al.} obtained a numerical solution which they parameterized as
\begin{equation} \label{eq:nubar0}
\bar{\nu}(\varepsilon) = \frac{\varepsilon}{1+kg(\varepsilon)},
\end{equation}
with $k$ defined by Eq. \ref{eq:dedr}. The function $g(\varepsilon)$ is merely plotted in \cite{Lindhard:1963a}, and a frequently used parameterization, 
\begin{equation}
g(\varepsilon) = 3\varepsilon ^{0.15} + 0.7\varepsilon ^{0.6} + \varepsilon,
\end{equation}
is given in \cite{Lewin:1996}. In the context of direct detection of dark matter, one is usually interested in the fraction of energy given to electrons (``the quenching factor''), 
\begin{equation} \label{eq:quenching}
f_n \equiv \frac{\varepsilon-\bar{\nu}}{\varepsilon} = \frac{kg(\varepsilon)}{1+kg(\varepsilon)},
\end{equation}
in which the subscript indicates that this fraction is for nuclear recoils.

\subsection{Trust, but verify} \label{ss:tbv}
As a preamble to studying solutions at low energy, the validity of Eq. \ref{eq:nubar0} was first verified for the case of germanium. Immediately, one must contend with the fact that Lindhard et al. did not calculate $f(t^{1/2})$ for $\varepsilon<0.002$. The authors of \cite{Wilson:1977} showed that errors in the nuclear stopping potential can be reduced from $>100\%$ to $<10\%$ using a Moli\`ere parameterization of the screening. For $\varepsilon<0.002$, I therefore extended the $f(t^{1/2})$ given in Fig. 2 of \cite{Lindhard:1963a} using this parameterization, Eq. 15 of \cite{Wilson:1977}. The transition is smooth and continuous. Note that the $S_n(\varepsilon)$ given therein may be transformed to the desired $f(t^{1/2})$ by differentiation of $\varepsilon S_n(\varepsilon)$.

The point of this choice is to preserve the original treatment of \cite{Lindhard:1963a} to the greatest extent possible, while obtaining a reasonable form of the nuclear potential for low energies. This is not the only reasonable choice, and in Sec.~\ref{ss:up}, I discuss the effect of the so-called universal potential \cite{Ziegler:1985}.

\begin{figure}[ht]
\begin{center}
%\vskip -0.4cm
\includegraphics[width=0.48\textwidth]{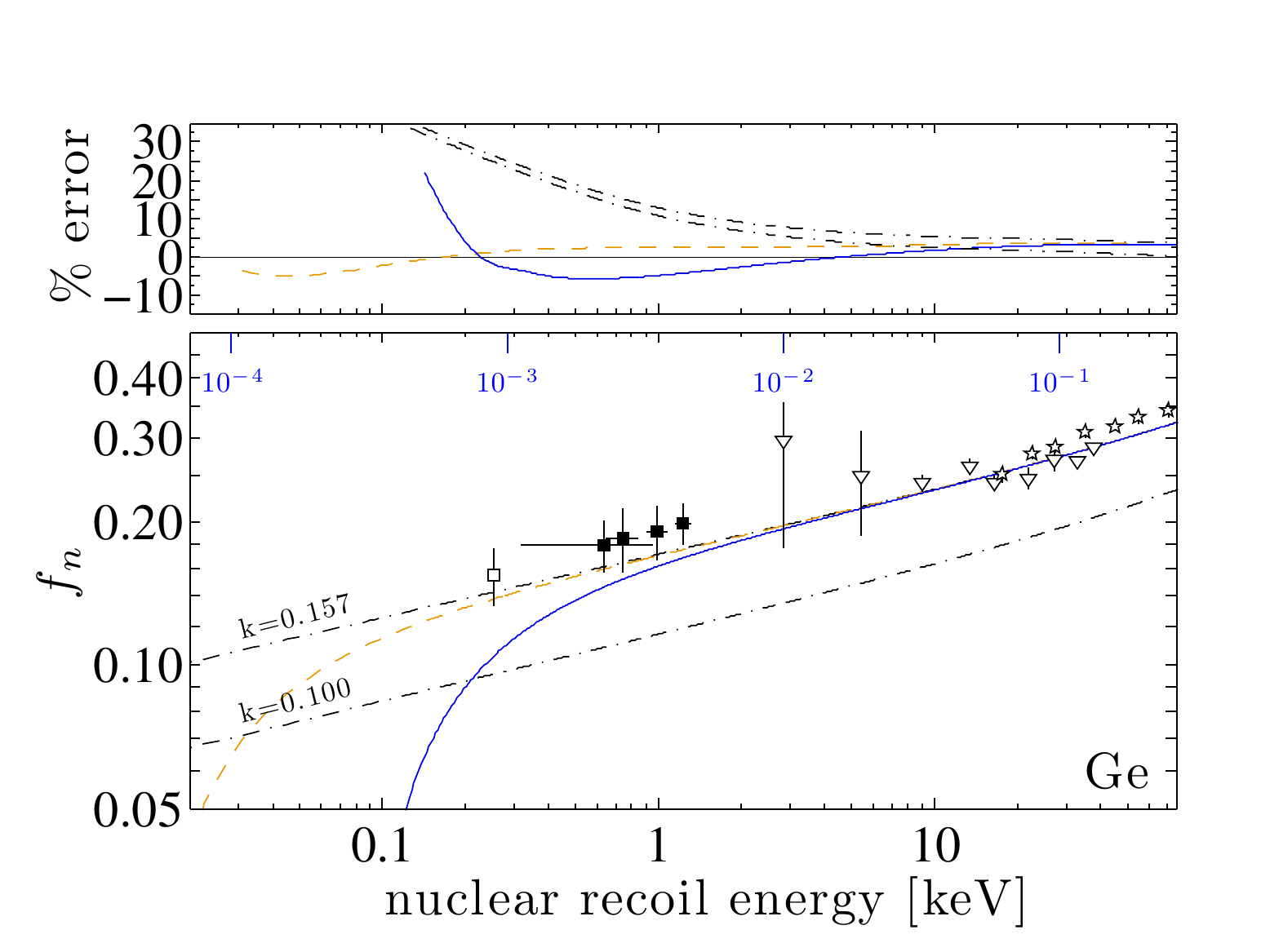}
%\vskip -0.1cm
\caption{{\bf (upper panel)} For germanium, percent error in the solution of Eq. \ref{eq:idiff} and Eq.~$1'$, as defined in the text. {\bf (lower panel)} Solutions of Eq. \ref{eq:idiff} (dash-dot: Eq.~\ref{eq:nubar0}, and dashed: Eq.~\ref{eq:nubar1}) and Eq.~$1'$ (solid). Note that Eq.~$1'$ corresponds to $u>0$, as discussed in Sec. \ref{ss:ab}. The inset scale indicates values of $\varepsilon$. See also Table \ref{table1}. Data are from \cite{Jones:1975, Barbeau:2007,Messous:1995,Shutt:1992}.}
%\vskip -0.5cm
\label{fig1}
\end{center}
\end{figure}

The residual fractional error in solutions of Eq. \ref{eq:idiff} was assessed by the quotient $(lhs-rhs)/(lhs+rhs)$, where $lhs$ and $rhs$ refer to the left and right hand sides of Eq. \ref{eq:idiff}. This is plotted as a percent in Fig. \ref{fig1}, above the standard solution (dash-dot curves). One can now see an additional source of uncertainty in the standard solution (Eq. \ref{eq:nubar0}) at low energies: the residual error grows from $<5\%$ at $\varepsilon=10^{-1}$ to about 25\% by $\varepsilon=10^{-3}$ (and, not shown, more than 50\% by $\varepsilon=10^{-4}$).

\section{Solutions near threshold} \label{sec:le}
This section focuses on numerical solutions of Eq. \ref{eq:idiff} near the energy threshold of ionization (or scintillation plus ionization) detectors. In practice this means the energy range from a few tens of eV up to a few keV.  A solution at any particular value of $\varepsilon$ requires knowledge of $\bar{\nu}(\varepsilon)$ at all smaller values of $\varepsilon$, so it is helpful to find a simple parameterization. While a simpler power law provides a good solution near threshold, it seems preferable to modify the standard solution to provide smaller error and more flexibility. This is accomplished by adding a constant $q$ to Eq. \ref{eq:nubar0}:
\begin{equation}\label{eq:nubar1}
\bar{\nu}(\varepsilon) = \frac{\varepsilon}{1+kg(\varepsilon)} + q.
\end{equation}
A sufficiently small $q$ has almost no effect on the solution for $E\gtrsim 1$~keV, yet for smaller energies allows a sharp cutoff (or enhancement, if $q$ were negative) in the energy given to electrons. This becomes particularly important in the absence of approximation (B).

The result of solving Eq. \ref{eq:idiff} with Eq. \ref{eq:nubar1} for germanium is shown in Fig. \ref{fig1} (dashed curve), with the parameter value given in Table \ref{table1}. The quenching prediction is then given by
\begin{equation} \label{eq:qp}
f_n = \frac{kg(\varepsilon)}{1+kg(\varepsilon)} - q/\varepsilon,
\end{equation}
which begins to differ from the standard solution (Eq. \ref{eq:nubar0}) below about one hundred eV. At the same time, the residual error in the solution is reduced to a few percent across the entire range of $\varepsilon$ in question. 

\subsection{Past steps and next steps}
The remainder of this section will explore the effects of approximation (B). Prior to this it is worth making a few comments about previous work. 

Recent calculations by Barker and Mei \cite{Barker:2012ek} examine the Lindhard model, with specific attention to ionization effects due to nuclear scattering. Essentially, this can be thought of as questioning approximation (D). Their results show a large decrease in $f_n$ above a few tens of keV, and almost no change at  1~keV. Their calculations do not extend below 1~keV.

A comparison between the Lindhard model and the widely used SRIM code (described in \cite{Ziegler:1985}) can be seen in \cite{Mangiarotti:2006ye}. This work highlights an important divide: The SRIM code uses the universal potential \cite{Ziegler:1985} for the nuclear scattering, which is evidently the most accurate potential for the widest selection of pairs of nuclei. While the SRIM code is widely used, its predictions at low energy do not agree particularly well with available data (as shown in \cite{Mangiarotti:2006ye}). It is not known if this is due to the choice of potential (probably not) or rather to the implementation of the energy loss calculations. The code itself is not available to scrutinize nor modify, so it is of interest to see how the universal potential modifies the Lindhard model. This point has been made previously in a slightly different context \cite{Bezrukov:2010qa}. These effects are discussed in Sec. \ref{ss:up}.

\subsection{Atomic binding energy approximation} \label{ss:ab}
In the slowing down of an recoil atom, some of the energy that is given to electrons must be spent on atomic binding. The Lindhard model considers average quantities, so the relevant binding energy in this context is arguably the solid state average energy required to produce an electron-hole pair. This quantity is well known in germanium to have a value of 3.0 eV \cite{Agnese:2013jaa}. In the reduced units defined by Eq. \ref{eq:epsilon}, the value is $u = 1.06\times10^{-5}$. 

It can be shown that approximation (B) is removed from the original Lindhard model by replacing the term $\bar{\nu}(t/\varepsilon)$ in Eq. \ref{eq:idiff} with $\bar{\nu}(t/\varepsilon-u)$. I will refer to Eq. \ref{eq:idiff} with this modification as Eq.~$1'$. The result of solving Eq.~$1'$ with Eq. \ref{eq:nubar1} is shown in Fig. \ref{fig1} (solid curve). The primary effect is a fairly sharp cutoff in the fraction of energy given to electrons, at a nuclear recoil energy $E\sim100$~eV. The \% error in the solution increases rapidly at the cutoff point, because the derivative $\bar{\nu}'$ approaches a constant value (and cannot keep pace with the decline in $\bar{\nu}$). For the sake of completeness, note that forcing $q=0$ (no cutoff) results in a significantly larger error.

\subsection{Other target media} \label{ss:otm}
\begin{figure}[ht]
\begin{center}
%\vskip -0.4cm
\includegraphics[width=0.48\textwidth]{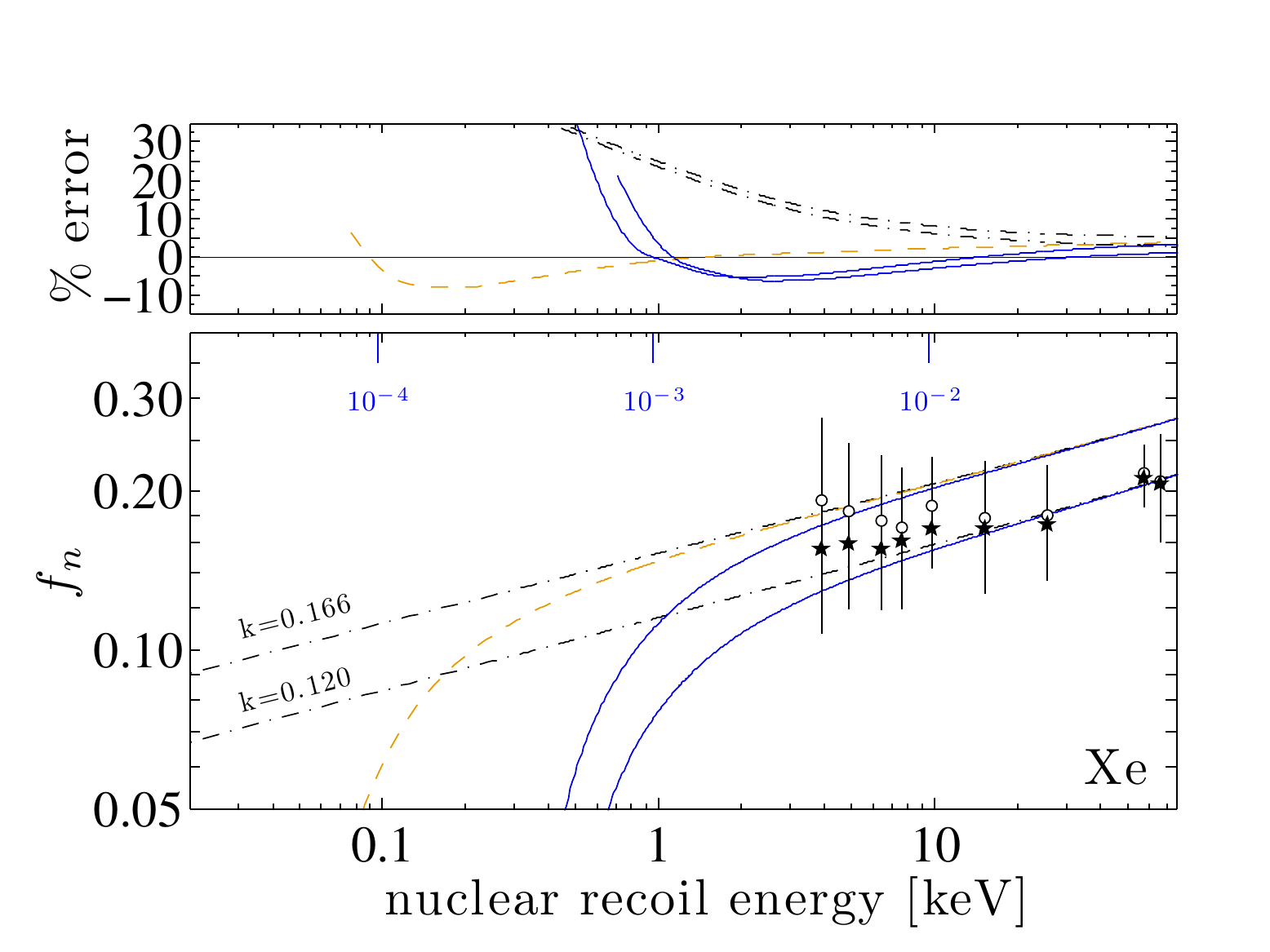}
%\vskip -0.1cm
\caption{Curves as described in Fig. \ref{fig1}, here for xenon. Also shown is a solution of Eq.~$1'$ with non-standard $k=0.120$. Data are from \cite{Manzur:2009hp}, and with threshold correction from \cite{Sorensen:2011bd}.}
%\vskip -0.5cm
\label{fig2}
\end{center}
\end{figure}

\begin{figure}[ht]
\begin{center}
%\vskip -0.4cm
\includegraphics[width=0.48\textwidth]{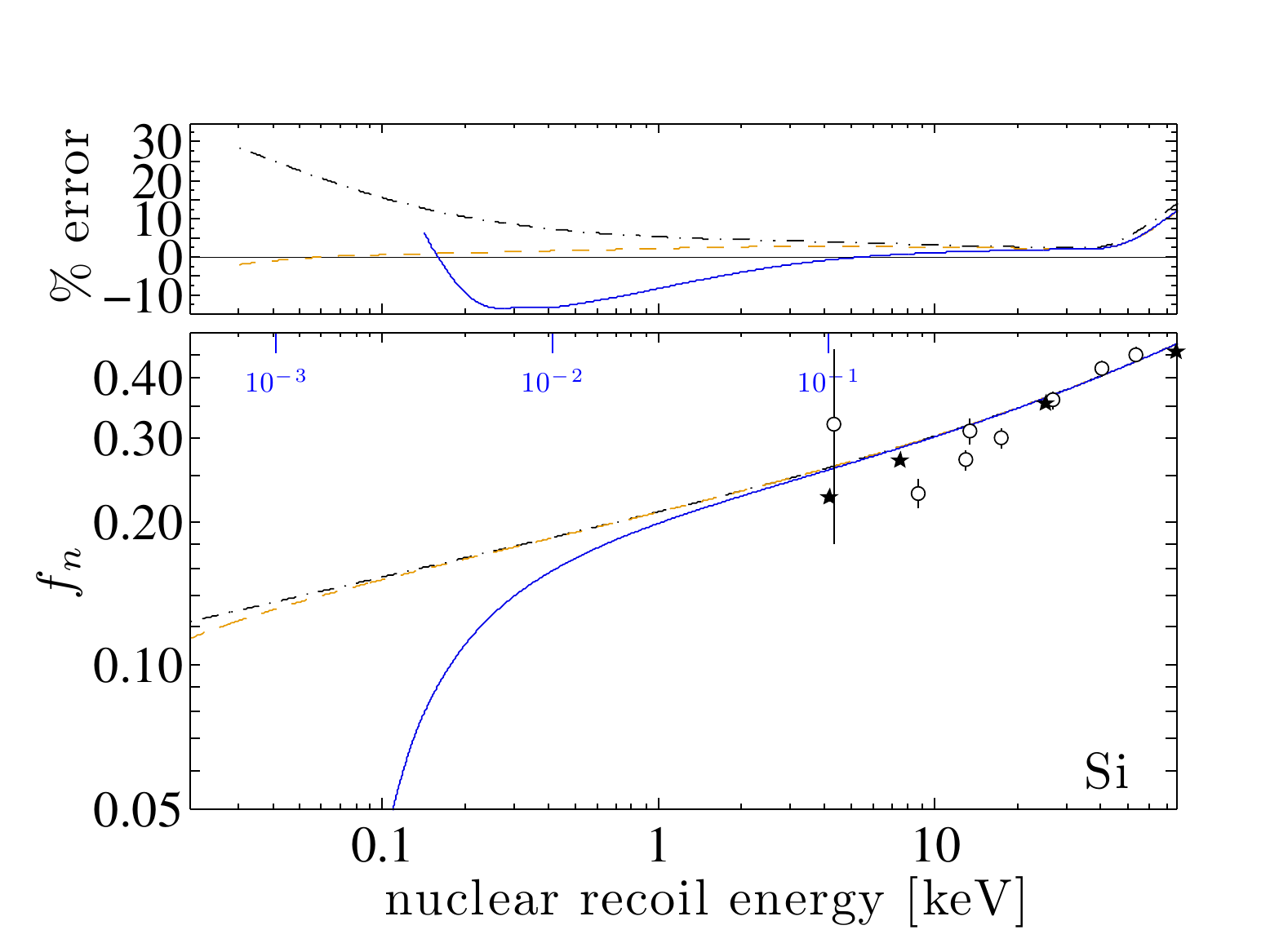}
%\vskip -0.1cm
\caption{Curves as described in Fig. \ref{fig1}, here for silicon. Data are from \cite{Zecher:1990,Dougherty:1992}.}
%\vskip -0.5cm
\label{fig3}
\end{center}
\end{figure}

\begin{figure}[ht]
\begin{center}
%\vskip -0.4cm
\includegraphics[width=0.48\textwidth]{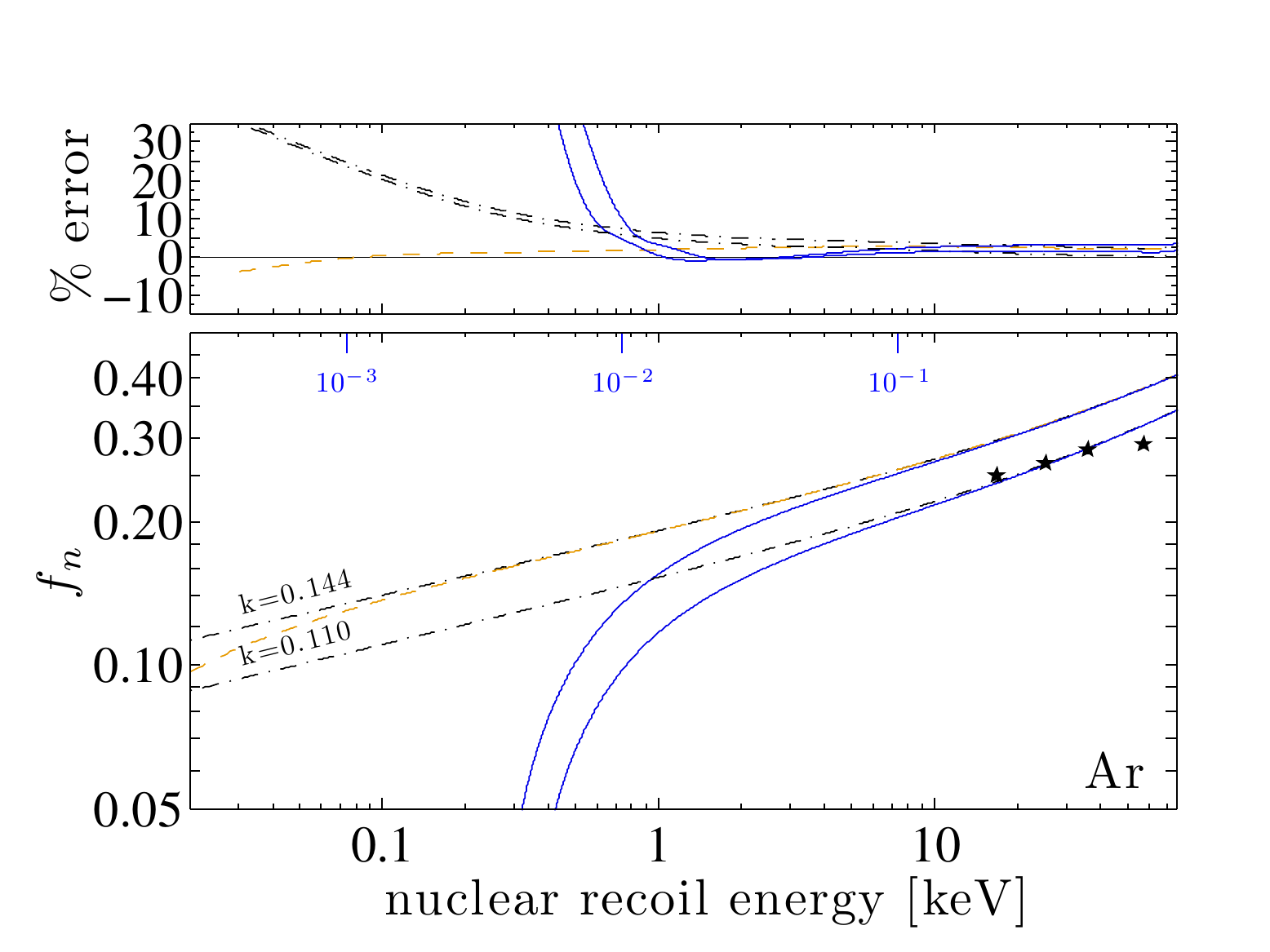}
%\vskip -0.1cm
\caption{Curves as described in Fig. \ref{fig1}, here for argon. Also shown is a solution of Eq.~$1'$ with non-standard $k=0.110$. Data are from \cite{Cao:2014gns}.}
%\vskip -0.5cm
\label{fig4}
\end{center}
\end{figure}

Several other target materials are of present interest in the search for a direct detection of dark matter. The treatment and conclusions for these materials are not qualitatively different from those obtained for germanium. Results are shown in Fig. \ref{fig2}, Fig. \ref{fig3} and Fig. \ref{fig4}, and summarized in Table \ref{table1}. For silicon, $u$ was taken to be the average energy 3.84~eV \cite{Cabrera:2010kh} required to create an electron hole pair. For argon and xenon, $u$ was taken to be the average energy required to create a single quanta (electron or photon). The values are 19.5~eV \cite{Doke:2002} and 13.8~eV \cite{Shutt:2007zz}.

The noble gases argon and xenon require additional comment. Data for these materials are consistent with a smaller total fraction of energy given to electrons than would be expected on the basis of Eq. \ref{eq:k}. This is not particularly troubling, considering the variety in calculations of $k$. As already mentioned, a solution of Eq. \ref{eq:idiff} at any particular value of $\varepsilon$ requires knowledge of $\bar{\nu}(\varepsilon)$ at all smaller values of $\varepsilon$, and this suggests that higher energy data must provide a normalization for the quenching prediction. Such a normalization is shown in Fig. \ref{fig2} and Fig. \ref{fig4}, where the 
values of $k$ were chosen by eye to approximately follow data for energies with $E\gtrsim10$~keV. This approach is reasonable in the absence of significant deviation from velocity proportional stopping, as discussed in Sec. \ref{ss:elec}.

\subsection{The universal potential} \label{ss:up}
The low-energy behavior of the Thomas-Fermi nuclear potential, and in particular this choice of screening function, has long been cited as a weakness of the Lindhard model. In fact, it is simply an input to the model. Many different approximations and parameterizations exist. As mentioned above, the Moli\`ere potential has been shown to be an excellent choice for a variety of atom pairs \cite{Wilson:1977}. %The effects of this potential for large impact parameter scattering (small $\varepsilon$) have already already been demonstrated ({\it cf.} Fig. \ref{fig1}). 

In order to understand the effect of the choice of potential, I solved Eq. \ref{eq:idiff} with $f(t^{1/2})$ corresponding to the universal potential \cite{Ziegler:1985}. The universal potential is just the same Coulomb potential discussed in Sec.~\ref{ss:nucl}, with a slightly weaker screening function $\phi_0(r/a)$ and a different length scale, $a=0.8853a_0/(2Z)^{0.23}$.

To ensure that the normalization of the solution was not constrained, I first used a simple power law parameterization $\bar{\nu}= \varepsilon - c \varepsilon^{1.18} + q$, in which $c$ is a constant. This function gives a good fit over the range shown in the figures, with error similar to the cases already discussed.  In the absence of approximation (B), the cutoff due to atomic binding tends to occur at an energy which is higher by about a factor $\times 2$.  

Based on this, it seemed preferable to use Eq.~\ref{eq:nubar0} to define $\bar{\nu}$ in this case as well. The resulting cutoff parameter does not depend on this choice to any significant degree. The error in the solution does increase slightly, to an average of typically 5\% over the range of interest \footnote{at energies $E\gtrsim100$~keV, the error increases more significantly. This may indicate the need for a better high-$E$ normalization, which is outside the scope the present work.}. This is a reasonable penalty compared with the resulting simplicity of parameterizing all the solutions in the same manner. The increase in cutoff energy appears to result primarily from the factor of about $\times2$ decrease in the length scale $a$.

\setlength{\tabcolsep}{7pt}
\begin{table}[h]
\centering
\caption{Values of the cutoff parameter $q$ for solutions of Eq.~$1'$ (as described in Sec. \ref{ss:ab}) with Eq. \ref{eq:nubar1}. Values of $q$ and $u$ are shown multiplied by a factor $ \times 10^{5}$. The standard Lindhard model, characterized by $q=u=0$, is shown for comparison. Values of $k$ were calculated from Eq.~\ref{eq:k} unless noted$^*$. The nuclear potential is either Thomas-Fermi (TF) or the universal potential of Ziegler (ZU). $\bar{\theta}$ refers to the predicted average nuclear recoil energy (in eV) to ionize a single electron.}
\begin{tabular}{clllll}
\\
\toprule
Atoms & $q$ & $u$ & $k$ & $\phi_0$ & $\bar{\theta}$ \\
\hline
Si & 0 & 0 & 0.146 & TF & 31\\ 
Si & 0.46 & 0 & 0.146 & TF & 32\\ 
Si & 27.9 & 9.35 & 0.146 & TF & 101 \\ 
Si & 40.1 & 9.35 & 0.146 & ZU & 140 \\ 
\hline
Ar & 0 & 0 & 0.144 & TF & 253\\ 
Ar & 0.44 & 0 & 0.144 & TF & 254\\ 
Ar & 49.4 & 26.4 & 0.144 & TF & 447 \\ 
Ar & 48.8 & 26.4 & 0.110$^*$ & TF & 544\\ 
Ar & 73.4 & 26.4 & 0.110$^*$ & ZU & 676\\ 
\hline
Ge & 0 & 0 & 0.157 & TF & 29\\ 
Ge & 0.42 & 0 & 0.157 & TF & 38\\ 
Ge & 3.39 & 1.06 & 0.157 & TF & 101\\ 
Ge & 8.84 & 1.06 & 0.157 & ZU & 201\\ 
\hline
Xe & 0 & 0 & 0.166 & TF & 224\\ 
Xe & 0.53 & 0 & 0.166 & TF & 260\\ 
Xe & 4.20 & 1.44 & 0.166 & TF & 492\\ 
Xe & 4.02 & 1.44 & 0.120$^*$ & TF & 618\\ 
Xe & 7.21 & 1.44 & 0.120$^*$ & ZU & 821\\ 

\botrule
%\multicolumn{3}{l}{$^a$ footnote}\\
\end{tabular}

\label{table1}
\end{table}

\section{Effect on dark matter direct detection sensitivity}
\begin{figure}[h]
\begin{center}
%\vskip -0.4cm
\includegraphics[width=0.48\textwidth]{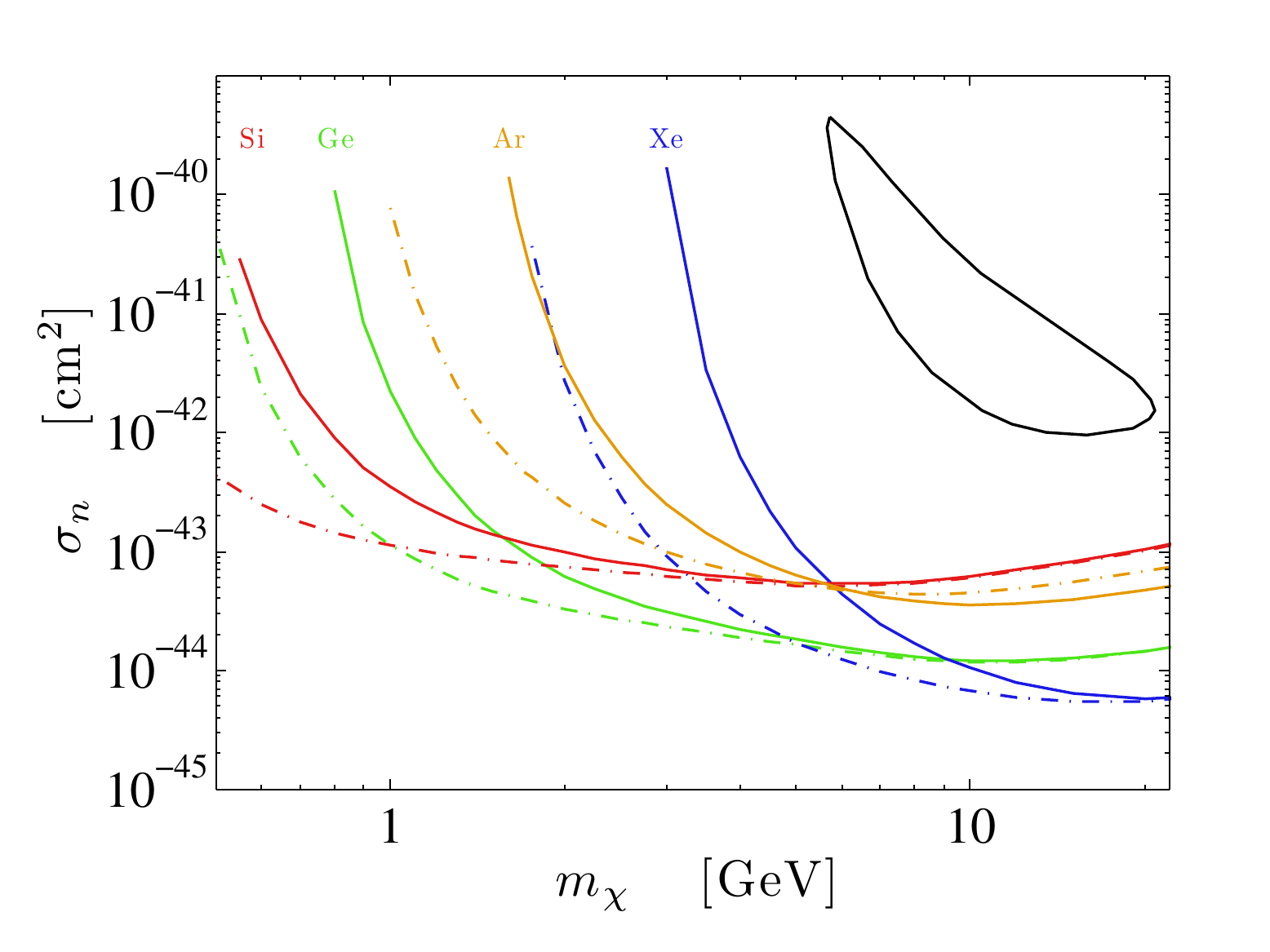}
%\vskip -0.1cm
\caption{Sensitivities of hypothetical 1000 kg-day exposure with a background rate of $10^{-3}$~counts/keV/kg/day and a search window from the threshold for ionization of a single electron up to 1 keV electron equivalent (unquenched) energy. As a landmark, a possible signal at CDMS \cite{Agnese:2013rvf} is indicated by the closed contour.  For each material, the dashed curve corresponds to the first line of Table \ref{table1}, and the solid curve to the last line.}
%\vskip -0.5cm
\label{fig5}
\end{center}
\end{figure}

In order to assess the effect of these results on the sensitivity of direct detection experiments, Fig. \ref{fig5} shows hypothetical exclusion limits for spin-independent elastic scattering with (solid curves) and without (dash-dot curves) the atomic binding approximation. The curves were generated using the maximum gap method \cite{Yellin:2002}. Assumptions include a 1000 kg-day exposure and a background-free search window from the threshold for ionization of a single electron up to 1 keV electron equivalent (unquenched) energy.  This corresponds to a background counting rate of approximately $10^{-3}$~counts/keV/kg/day. The neutrino floor varies slightly with target \cite{Billard:2013qya} and is not shown for clarity. It attains a peak value at about $\sigma_n=5\times10^{-45}$~cm$^2$ at 6~GeV.

It is important to note that in liquid argon and xenon, electronic excitation in the sense of $f_n$ results in both ionization and scintillation. At very low energies in liquid xenon, it appears that nuclear recoil energy partitions nearly equally into these two channels \cite{Sorensen:2011bd}. In liquid argon, the lowest energy data point is consistent with equal partitioning. In calculating sensitivities, I have assumed the fraction is exactly 0.5 in both of these materials, and that only the ionization is measurable at very low energies. The magnitude of the applied electric field may change these assumptions somewhat.

The basic result of the atomic threshold cutoff in $f_n$ is a sharp decrease in sensitivity to low mass dark matter. The low mass region of parameter space is compatible with several putative detections reported in the past few years. This is because in most of these cases, hints of signal have tended to appear near the detection threshold. While sorting out signal from noise (or background) near detector threshold is a separate problem, interpretation and comparison of results relies critically on knowing the actual energy equivalent of a detector's threshold.

\section{Summary}
This article has shown that a kinematic cutoff due to atomic binding energy is an inherent part of the widely used Lindhard model. This is particularly important because it is generally recognized that such a cutoff must exist, but no direct connection to the model existed. Most previous work has converted a measured detector response to nuclear recoil energy using the Lindhard model. The model validity has either been assumed to hold to the ionization threshold (e.g. superCDMS SNOLAB in \cite{Cushman:2013zza}, DAMIC \cite{Barreto:2011zu}), or an ad-hoc kinematic cutoff has been inserted (e.g. XENON10 \cite{Angle:2011th} and LUX \cite{Akerib:2013tjd}).

Other kinematic effects may exist. In particular, a non-zero $\varepsilon$-intercept to the electronic stopping, as discussed in Sec. \ref{ss:elec}, would compound the kinematic cutoff. Unfortunately, one can only speculate about the likelihood and magnitude of such an effect, due to a paucity of data. Perhaps the most relevant data to the present context are shown in Fig.~4 of \cite{Markin:2009}. This is because the SiO$_2$ target used therein is also composed of covalent bonds. Velocity proportional stopping for helium projectiles is extrapolated to zero velocity in that case, consistent with Eq.~\ref{eq:dedr}. This suggests that a similar response might be expected for the materials considered in this article.

A full quantum model of atomic projectile range and energy loss in a solid would be a welcome addition to the literature. A clue to the complexity of such a task may be found in the fact that the Lindhard model, as originally formulated, remains widely used after more than 50 years. In the short term, it is hoped that the present extension of that model will provide useful guidance for expectations of the quenching of very low energy nuclear recoils. Crucially, new low-energy measurements may be able to discern between the slow decrease predicted by the original model, and the sharp cutoff predicted in the present work. 

%discuss the fact that hartree fock solid state electron wave functions are non-zero for distances which exceed the average interatomic separation? SRIM uses the smooth approximation, with considerable success...

\section*{Acknowledgments}
The author gratefully acknowledges support from the U.S. Department of Energy (DOE), Office of Science, Office of High Energy Physics under contract No. DE-AC02-05CH11231. Discussions with Jeremy Mardon were particularly helpful.


\begin{thebibliography}{References}

\bibitem{Lin:2011gj} 
  T.~Lin, H.~B.~Yu and K.~M.~Zurek,
  %``On Symmetric and Asymmetric Light Dark Matter,''
  Phys.\ Rev.\ D {\bf 85}, 063503 (2012)
  [arXiv:1111.0293 [hep-ph]].
  %%CITATION = ARXIV:1111.0293;%%
  
\bibitem{Hooper:2012cw} 
  D.~Hooper, N.~Weiner and W.~Xue,
  %``Dark Forces and Light Dark Matter,''
  Phys.\ Rev.\ D {\bf 86}, 056009 (2012)
  [arXiv:1206.2929 [hep-ph]].
  %%CITATION = ARXIV:1206.2929;%%

\bibitem{Pyle:2012mz} 
  M.~Pyle, D.~A.~Bauer, B.~Cabrera, J.~Hall, R.~W.~Schnee, R.~B.~Thakur and S.~Yellin,
  %``Low-Mass WIMP Sensitivity and Statistical Discrimination of Electron and Nuclear Recoils by Varying Luke-Neganov Phonon Gain in Semiconductor Detectors,''
  J.\ Low.\ Temp.\ Phys.\  {\bf 167}, 1081 (2012)
  [arXiv:1201.3685 [astro-ph.IM]].
  %%CITATION = ARXIV:1201.3685;%%

\bibitem{Angle:2011th} 
  J.~Angle {\it et al.}  [XENON10 Collaboration],
  %``A search for light dark matter in XENON10 data,''
  Phys.\ Rev.\ Lett.\  {\bf 107}, 051301 (2011)
  [Erratum-ibid.\  {\bf 110}, 249901 (2013)]
  [arXiv:1104.3088 [astro-ph.CO]].
  %%CITATION = ARXIV:1104.3088;%%
    
\bibitem{Essig:2011nj} 
  R.~Essig, J.~Mardon and T.~Volansky,
  %``Direct Detection of Sub-GeV Dark Matter,''
  Phys.\ Rev.\ D {\bf 85}, 076007 (2012)
  [arXiv:1108.5383 [hep-ph]].
  %%CITATION = ARXIV:1108.5383;%%

\bibitem{Graham:2012su} 
  P.~W.~Graham, D.~E.~Kaplan, S.~Rajendran and M.~T.~Walters,
  %``Semiconductor Probes of Light Dark Matter,''
  Phys.\ Dark Univ.\  {\bf 1}, 32 (2012)
  [arXiv:1203.2531 [hep-ph]].
  %%CITATION = ARXIV:1203.2531;%%
  
\bibitem{Lindhard:1963a} J. Lindhard, V.~Nielsen,~M.~Scharff and P.V.~Thomsen, Mat. Fys. Medd. Dan. Vid. Selsk. {\bf33} 10 (1963).

\bibitem{Aalseth:2012if} 
  C.~E.~Aalseth {\it et al.}  [CoGeNT Collaboration],
  %``CoGeNT: A Search for Low-Mass Dark Matter using p-type Point Contact Germanium Detectors,''
  Phys.\ Rev.\ D {\bf 88}, no. 1, 012002 (2013)
  [arXiv:1208.5737 [astro-ph.CO]].
  %%CITATION = ARXIV:1208.5737;%%

\bibitem{Agnese:2013jaa} 
  R.~Agnese {\it et al.}  [SuperCDMS Collaboration],
  %``Search for Low-Mass Weakly Interacting Massive Particles Using Voltage-Assisted Calorimetric Ionization Detection in the SuperCDMS Experiment,''
  Phys.\ Rev.\ Lett.\  {\bf 112}, no. 4, 041302 (2014)
  [arXiv:1309.3259 [physics.ins-det]].
  %%CITATION = ARXIV:1309.3259;%%

\bibitem{Barreto:2011zu} 
  J.~Barreto {\it et al.}  [DAMIC Collaboration],
  %``Direct Search for Low Mass Dark Matter Particles with CCDs,''
  Phys.\ Lett.\ B {\bf 711}, 264 (2012)
  [arXiv:1105.5191 [astro-ph.IM]].
  %%CITATION = ARXIV:1105.5191;%%

\bibitem{Cushman:2013zza} 
  P.~Cushman, C.~Galbiati, D.~N.~McKinsey, H.~Robertson, T.~M.~P.~Tait, D.~Bauer, A.~Borgland and B.~Cabrera {\it et al.},
  %``Working Group Report: WIMP Dark Matter Direct Detection,''
  arXiv:1310.8327 [hep-ex].
  %%CITATION = ARXIV:1310.8327;%%

\bibitem{Lindhard:1963b} J. Lindhard,~M.~Scharff and H.E.~Schiott, Mat. Fys. Medd. Dan. Vid. Selsk. {\bf33} 14 (1963).

\bibitem{Lindhard:1968}  J. Lindhard,~V.~Nielsen and M.~Scharff, Mat. Fys. Medd. Dan. Vid. Selsk. {\bf36} 10 (1968).
  
\bibitem{Wilson:1977}
W.D.~Wilson, L.G.~Haggmark and J.P.~Biersack, Phys. Rev. B {\bf15} 2458 (1977).

\bibitem{Ziegler:1985}
J.F.~Ziegler, J.P.~Biersack and U.~Littmark, ``The Stopping and Ranges of Ions in Matter, Vol. 1,'' Pergamon Press, 1985.

\bibitem{Land:1977} D. J. Land, J. G. Brennan, . D. G, Simons, and M. D. Brown, Phys. Rev. A {\bf16} 492 (1977).

\bibitem{Moller:2002}
S. P. M\o ller {\it et al.}, Phys Rev. Lett. {\bf88} 193201 (2002).

\bibitem{Sigmund:2015}
P. Sigmund and A. Schinner, Nucl. Instr. Meth. B {\bf342} 292 (2015).

\bibitem{Markin:2009}
S. N. Markin, D. Primetzhofer, and P. Bauer, Phys Rev. Lett. {\bf103} 113201 (2009).

\bibitem{Ahlen:1983}
S.P.~Ahlen and G.~Tarl\'e, Phys. Rev. D {\bf27} 688 (1983).

\bibitem{Collar:2010gg} 
  J.~I.~Collar and D.~N.~McKinsey,
  %``Comments on 'First Dark Matter Results from the XENON100 Experiment',''
  arXiv:1005.0838 [astro-ph.CO].

\bibitem{Lewin:1996}
J.D.~Lewin and P.F.~Smith, Astropart. Phys. {\bf6} 87 (1996).

\bibitem{Jones:1975} K. W. Jones and H. W. Kraner, Phys. Rev. A {\bf11} 1347 (1975).

\bibitem{Barbeau:2007} P.S.~Barbeau, J.I.~Collar and O.~Tench, J. Cosmol. Astropart. Phys. {\bf09} 009 (2007).

\bibitem{Messous:1995} Y.~Messous {\it et al.}, Astropart. Phys. {\bf3} 361 (1995).

\bibitem{Shutt:1992} T.~Shutt {\it et al.}, Phys. Rev. Lett. {\bf69} 3425 (1992).

\bibitem{Barker:2012ek} 
  D.~Barker and D.~M.~Mei,
  %``Germanium Detector Response to Nuclear Recoils in Searching for Dark Matter,''
  Astropart.\ Phys.\  {\bf 38}, 1 (2012)
  [arXiv:1203.4620 [astro-ph.IM]].
  %%CITATION = ARXIV:1203.4620;%%
  
\bibitem{Mangiarotti:2006ye} 
  A.~Mangiarotti, M.~I.~Lopes, M.~L.~Benabderrahmane, V.~Chepel, A.~Lindote, J.~Pinto da Cunha and P.~Sona,
  %``A Survey of energy loss calculations for heavy ions between 1-keV and 100-keV,''
  Nucl.\ Instrum.\ Meth.\ A {\bf 580}, 114 (2007)
  [physics/0610286].

\bibitem{Bezrukov:2010qa} 
  F.~Bezrukov, F.~Kahlhoefer, M.~Lindner, F.~Kahlhoefer and M.~Lindner,
  %``Interplay between scintillation and ionization in liquid xenon Dark Matter searches,''
  Astropart.\ Phys.\  {\bf 35}, 119 (2011)
  [arXiv:1011.3990 [astro-ph.IM]].
  %%CITATION = ARXIV:1011.3990;%%

\bibitem{Cabrera:2010kh} 
  B.~Cabrera, M.~Pyle, R.~Moffatt, K.~Sundqvist and B.~Sadoulet,
  %``Oblique propagation of electrons in crystals of germanium and silicon at sub-Kelvin temperature in low electric fields,''
  arXiv:1004.1233 [astro-ph.IM].
  %%CITATION = ARXIV:1004.1233;%%.

\bibitem{Doke:2002} T.~Doke {\it et al.}, Jpn. J. Appl. Phys.  {\bf41} 1538 (2002).

\bibitem{Shutt:2007zz} 
  T.~Shutt, A.~Bolozdynya, P.~Brusov, C.~E.~Dahl and J.~Kwong,
  %``Performance and fundamental processes at low energy in a two-phase liquid xenon dark matter detector,''
  Nucl.\ Phys.\ Proc.\ Suppl.\  {\bf 173}, 160 (2007).
  %%CITATION = NUPHZ,173,160;%%

\bibitem{Manzur:2009hp} 
  A.~Manzur, A.~Curioni, L.~Kastens, D.~N.~McKinsey, K.~Ni and T.~Wongjirad,
  %``Scintillation efficiency and ionization yield of liquid xenon for mono-energetic nuclear recoils down to 4 keV,''
  Phys.\ Rev.\ C {\bf 81}, 025808 (2010)
  [arXiv:0909.1063 [physics.ins-det]].
  %%CITATION = ARXIV:0909.1063;%%

\bibitem{Sorensen:2011bd} 
  P.~Sorensen and C.~E.~Dahl,
  %``Nuclear recoil energy scale in liquid xenon with application to the direct detection of dark matter,''
  Phys.\ Rev.\ D {\bf 83}, 063501 (2011)
  [arXiv:1101.6080 [astro-ph.IM]].
  %%CITATION = ARXIV:1101.6080;%%
    
\bibitem{Zecher:1990} P. Zecher, D. Wang, J. Rapaport, C. J. Martoff and B.A. Young, Phys Rev. A {\bf41} 4058 (1990).

\bibitem{Dougherty:1992} B.L.~Dougherty, Phys. Rev. A {\bf45} 2104 (1992).

\bibitem{Cao:2014gns} 
  H.~Cao {\it et al.}  [SCENE Collaboration],
  %``Measurement of Scintillation and Ionization Yield and Scintillation Pulse Shape from Nuclear Recoils in Liquid Argon,''
  arXiv:1406.4825 [physics.ins-det].
  %%CITATION = ARXIV:1406.4825;%%

\bibitem{Yellin:2002} S.~Yellin, Phys. Rev. D {\bf66} 032005 (2002).

\bibitem{Billard:2013qya} 
  J.~Billard, L.~Strigari and E.~Figueroa-Feliciano,
  %``Implication of neutrino backgrounds on the reach of next generation dark matter direct detection experiments,''
  Phys.\ Rev.\ D {\bf 89}, 023524 (2014)
  [arXiv:1307.5458 [hep-ph]].
  %%CITATION = ARXIV:1307.5458;%%

\bibitem{Agnese:2013rvf} 
  R.~Agnese {\it et al.}  [CDMS Collaboration],
  %``Silicon Detector Dark Matter Results from the Final Exposure of CDMS II,''
  Phys.\ Rev.\ Lett.\  {\bf 111}, 251301 (2013)
  [arXiv:1304.4279 [hep-ex]].
  %%CITATION = ARXIV:1304.4279;%%

\bibitem{Akerib:2013tjd} 
  D.~S.~Akerib {\it et al.}  [LUX Collaboration],
  %``First results from the LUX dark matter experiment at the Sanford Underground Research Facility,''
  Phys.\ Rev.\ Lett.\  {\bf 112}, no. 9, 091303 (2014)
  [arXiv:1310.8214 [astro-ph.CO]].
  %%CITATION = ARXIV:1310.8214;%%
    
\end{thebibliography}
\end{document}